\documentclass[prl,twocolumn,groupedaddress,showpacs]{revtex4b4}
\usepackage{graphicx}
\begin{document}
\title{Entropic effects on the Size Evolution of Cluster Structure}
\author{Jonathan P.~K.~Doye}
\affiliation{University Chemical Laboratory, Lensfield Road, Cambridge CB2 1EW, United Kingdom}
\author{Florent Calvo}
\thanks{Permanent address: Laboratoire de Physique Quantique,
IRSAMC, Universit\'e Paul Sabatier, 118 Route de Narbonne, F31062 Toulouse Cedex.} 
\affiliation{University Chemical Laboratory, Lensfield Road, Cambridge CB2 1EW, United Kingdom}
\begin{abstract}
We show that the vibrational entropy can play a crucial role in 
determining the equilibrium structure of clusters by 
constructing structural phase diagrams showing how the 
structure depends upon both size and temperature. 
These phase diagrams are obtained for example rare gas and metal clusters.
\end{abstract}
\maketitle

\begin{figure}
\begin{center}
\includegraphics[width=8.2cm]{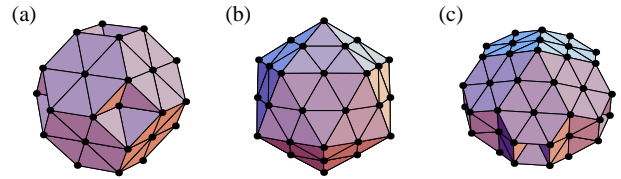}
\vglue -0.2cm
\begin{minipage}{8.5cm}
\caption{\label{fig:forms}
Three examples of the structures clusters can adopt:
(a) a fcc 38-atom truncated octahedron,
(b) a 55-atom Mackay icosahedron \cite{Mackay}, and (c) a 75-atom Marks decahedron \cite{Marks84}.
These clusters have the optimal shape for the three main types of 
regular packing seen in clusters: face-centred cubic, icosahedral and decahedral, respectively.
The latter two structural types cannot be extended to bulk because of the five-fold axes of symmetry.
}
\end{minipage}
\end{center}
\end{figure}

Much of the interest in clusters or nanoparticles derives from the insights they
can provide into how properties emerge and evolve on going between
the atomic and molecular and bulk limits. 
Cluster structure provides a particular interesting example of this size evolution.
At large enough sizes the clusters must display the bulk crystalline structure,
but this limit may sometimes only be achieved at very large sizes (e.g. at least
$20\,000$ atoms for sodium clusters \cite{Martin90}) 
and before that limit is reached unusual structural forms are often observed. 
For example, many clusters bound by van der Waals or metallic forces 
exhibit structures with five-fold axes of symmetry, a possibility that 
is forbidden in bulk crystalline materials.
For these clusters the dominant structural motif typically
changes from icosahedral (Fig.~\ref{fig:forms}(b)) to decahedral 
(Fig.~\ref{fig:forms}(c)) to face-centred-cubic (fcc) (Fig.~\ref{fig:forms}(a)) as the size increases. 

For many materials these structural changes occur at sizes that are 
too large for global optimization to be feasible.
Therefore, the typical theoretical approach to systematically investigating the
size evolution of cluster structure is to compare the energies of stable 
sequences of structures, such as the forms shown in Fig.~\ref{fig:forms}.
`Crossover sizes' are then identified where the sequence with lowest energy changes. 
At this crossover the most common equilibrium structure is expected to change.
This technique has been applied to rare gas \cite{Xie,van89,Raoult89a},
metal \cite{Cleveland91,Lim,Cleveland97b,Barreteau00} and  molecular clusters \cite{Calvo99a}.

The above approach is certainly valid at zero temperature, since 
the equilibrium structure then corresponds to the one with lowest energy.
At other temperatures, however, the structure with lowest free energy needs to be found. 
However, perhaps through an expectation that entropic effects are 
unlikely to be important or are too complicated to take into account, 
size is usually the only variable that 
is considered both experimentally \cite{Farges88,Kakar97,Torchet96} and theoretically 
\cite{Xie,van89,Raoult89a,Cleveland91,Lim,Cleveland97b,Barreteau00}.

In this paper we consider the role that entropy plays in the size
evolution of cluster structure, and show that
temperature can be a key variable in determining the equilibrium 
structure of a cluster.
A clue to this result can be garnered from
the growing number of examples
of solid-solid transitions in clusters where the structure changes from 
fcc or decahedral to icosahedral as the temperature increases
\cite{Doye95c,Doye98a,Doye99c,Cleveland98,BerryS00}.
The most well-investigated examples of these transitions are for those
small Lennard-Jones (LJ) clusters that have a non-icosahedral global minimum. 
For $N<150$ there are only 8 examples and these occur at sizes where the
non-icosahedral morphology has a near optimum shape, whereas the icosahedral 
structures have an incomplete outer layer. The global minimum is fcc at $N$=38, 
decahedral at $N$=75--77, 102--104 and 
has an unusual structure called a Leary tetrahedron at $N$=98 \cite{Leary99}.

Here we analyse these examples further in order
to identify 
what is the most significant contribution to the difference in entropies 
between the structural types, particularly as the size increases.
Direct computation of $T_{\rm ss}$, the temperature of the solid-solid transition,
is possible by parallel tempering \cite{Marinari92}, but this is computationally
demanding because of the large (free) energy barriers between the structural types \cite{Doye99c,Doye99f}, 
and has only been done for LJ$_{38}$ \cite{Neirotti00}.

The superposition method \cite{Wales93a,WalesDMMW00} provides another approach to 
this calculation and one which is particularly useful for our present purposes because 
it allows us to analyse the different contributions to the entropy. 
In this approach the partition function is written as a sum over all the minima 
on the potential energy surface, and by restricting the sum to a certain subset of the minima
the thermodynamic properties of a particular region of configuration space can be obtained. 
The centre of the solid-solid transition occurs when the partition functions
for the two competing structural types are equal, i.e.\ $Z_A=Z_B$.
In the harmonic approximation \cite{anharm} this then gives
\begin{equation}
\label{eq:HSM}
\sum_{i\in A} {n_i \exp(-\beta E_i)\over \overline{\nu}_i^{3N-6}}=
\sum_{j\in B} {n_j \exp(-\beta E_j)\over \overline{\nu}_j^{3N-6}},
\end{equation}
where $\beta$=$1/kT$, 
$E_i$ is the potential energy of minimum $i$, 
$\overline{\nu}_i$ is the geometric mean vibrational frequency, and
$n_i$=$2N!/h_i$ is the number of permutational isomers of $i$,
where $h_i$ is the order of the point group. 
$T_{\rm ss}$ values calculated using Eq.~(\ref{eq:HSM})
are given in Table \ref{table:Tss}. 
It is noteworthy that there is a general decrease in the values 
with increasing size.

\begin{table}
\caption{\label{table:Tss}Estimates of the solid-solid transition temperature $T_{\rm ss}$ 
for those Lennard-Jones clusters with less than 150 atoms that have non-icosahedral global minima.
$\overline{\nu}_A$ and $\overline{\nu}_B$ are the mean frequencies of the lowest-energy non-icosahedral
and icosahedral minima, respectively. $\epsilon$ is the equilibrium pair well depth of the LJ potential.
}
\begin{ruledtabular}
\begin{tabular}{cccccc}
 & \multicolumn{3}{c}{$T_{\rm ss}/\epsilon k^{-1}$} & \\
\cline{2-4} 
 $N$ & HSM & Einstein & Eq.\ (\ref{eq:t_ss_est}) & $\Delta E/\epsilon$ & $\overline{\nu}_A/\overline{\nu}_B$ \\
\hline
 38 & 0.121 & 0.199 & 0.316 & 0.676 & 1.0200 \\
 75 & 0.082 & 0.234 & 0.119 & 1.210 & 1.0475 \\
 76 & 0.046 & 0.223 & 0.053 & 0.510 & 1.0446 \\
 77 & 0.048 & 0.199 & 0.057 & 0.565 & 1.0451 \\
 98 & 0.004 & 0.009 & 0.006 & 0.022 & 1.0135 \\
102 & 0.013 & 0.096 & 0.014 & 0.086 & 1.0201 \\
103 & 0.016 & 0.116 & 0.018 & 0.107 & 1.0204 \\
104 & 0.007 & 0.069 & 0.008 & 0.048 & 1.0212 \\
\end{tabular}
\end{ruledtabular}
\end{table}

One contribution to the entropy of a morphology comes from the number of low-energy minima
of that type. 
For the current examples 
this term always favours the icosahedra because
there are many low-energy icosahedral minima with different arrangements
of the atoms in the incomplete outer layer \cite{Doye99f}.
Another entropic contribution can come from the symmetry of the cluster. 
At $N$=38, 75 and 98 the global minimum has high symmetry, thus reducing the
number of permutational isomers. For these three sizes this factor again favours 
the icosahedra. Lastly, there is the vibrational entropy. For the current examples
the icosahedral structures have a smaller mean vibrational frequency, which
again favours the icosahedra. 
However, unlike the previous two contributions to the entropy, 
this term favours the icosahedra for a LJ cluster of any size.
Furthermore, the absence of any solid-solid transitions when the LJ global minimum is icosahedral
therefore suggests that the vibrational entropy is crucial.

We can analyse further the effect of the vibrational entropy by applying 
an Einstein approximation, i.e.\ we assume that all the minima have the same 
mean vibrational frequency. The resulting values for $T_{\rm ss}$ are also 
given in Table \ref{table:Tss}. Although the transitions still occur, they do so
at significantly higher temperature with the error increasing with size,
because $\overline{\nu}$ is raised to the power $3N-6$ in Eq.~(\ref{eq:HSM}).
This increasing dominance of the vibrational entropy 
is the main reason that the actual $T_{\rm ss}$ generally decrease with size.
These results, therefore, suggest that solid-solid transitions, rather than being unusual,
should be expected for systems where different structural types have
systematic differences in $\overline{\nu}$ at sizes where the morphology with lower vibrational
entropy is the lowest in energy.

To obtain our results for $T_{\rm ss}$ we systematically searched the low-energy
regions of the potential energy surface for these clusters in order to generate the relevant 
samples of low-energy minima. However, this is not a practical approach at 
large sizes, therefore we seek a simpler way to estimate $T_{\rm ss}$.
Firstly, if we assume that all the minima associated with a morphology have
the same energy and vibrational frequency, then
\begin{equation}
T_{\rm ss}={\Delta E\over k \left(\log\left(n_B/n_A\right)+
         (3N-6)\log\left(\overline{\nu}_A/\overline{\nu}_B\right)\right)},
\end{equation}
where $\Delta E$=$E_B-E_A$ and $n_A$ is the total number of minima (geometric
and permutational isomers) associated with morphology A. 
Although in all the specific examples in Table \ref{table:Tss} 
there are far more icosahedral low-energy minima, 
on average the number of minima will be approximately the same.
Furthermore, at large sizes the vibrational term will dominate the denominator.
In this limit
\begin{equation}
\label{eq:t_ss_est}
T_{\rm ss}={\Delta E\over k (3N-6)\log\left(\overline{\nu}_A/\overline{\nu}_B\right)}
\end{equation}
We tested this expression for the examples in Table \ref{table:Tss} using the properties 
of the lowest-energy minima of the two competing morphologies. As expected this estimate
becomes more accurate as the size increases, and the errors for the largest sizes are small.

\begin{figure}
\includegraphics[width=8.2cm]{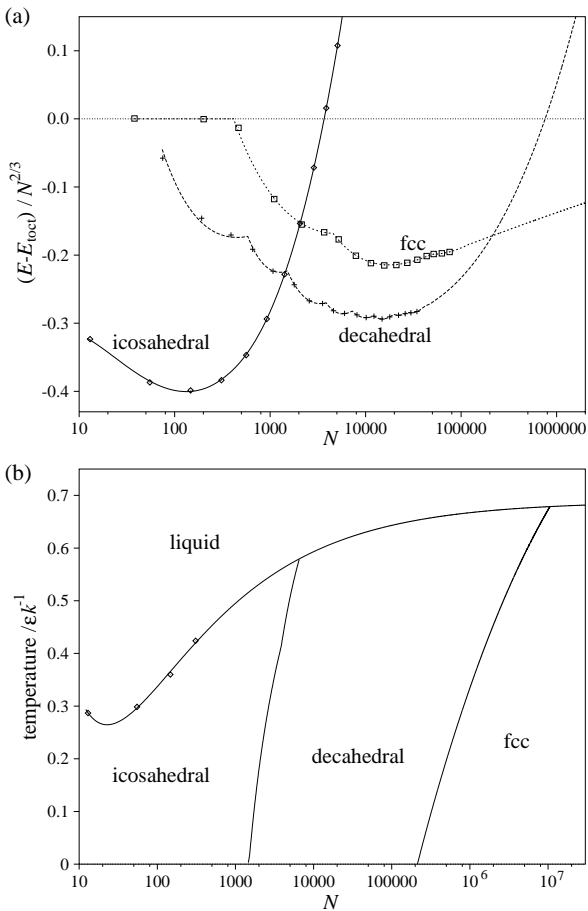}
\caption{\label{fig:LJphased}
(a) Energies of the competing structural types for LJ clusters.
The data points represent clusters with the optimal shape at that size, and the continuous lines
are fits to these data using Eq.~(\ref{eq:EvN}). 
The energy zero is $E_{\rm toct}$, a fit to the fcc truncated octahedra with regular hexagonal
$\{111\}$ faces (Fig.~\ref{fig:forms}(a)).
(b) Structural phase diagram for LJ clusters.
The data points represent the melting temperatures of 
the four smallest Mackay icosahedra. For argon $1\epsilon k^{-1}$$\equiv$121K.}
\end{figure}

We can also use Eq.~(\ref{eq:t_ss_est})
to determine how crossover sizes depend on temperature.
As with the theoretical studies mentioned earlier, we can compare stable sequences of
structures, but now monitoring not only their energies but also their vibrational frequencies.
These properties can be fitted to the forms
\begin{eqnarray}
\label{eq:EvN}
E&=&a_E N +b_E N^{2/3}+ c_E N^{1/3} + d_E \\
\label{eq:nuvN}
\overline{\nu}&=&a_{\nu} +{b_{\nu}\over N^{1/3}}+ {c_{\nu}\over N^{2/3}} + {d_{\nu}\over N}
\end{eqnarray}
where the first two terms represent volume and surface contributions, respectively.
These expressions can then be input into
Eq.~(\ref{eq:t_ss_est}) to map out the solid-solid transitions in the structural phase 
diagram. Finally to complete the diagram, the melting line has to be determined.
Again it will have the form 
\begin{equation}
\label{eq:Tm}
T_m=a_m +{b_m\over N^{1/3}}+ {c_m\over N^{2/3}} + {d_m\over N}.
\end{equation}

We first present results for LJ clusters, which provide a reasonable model of heavy
rare gas clusters, such as argon. Up to $N$$\approx$$10\,000$ the energies of 
the stable sequences of structures were obtained by minimization of all degrees of freedom.
Above this size minimizations were possible up to $N$$\approx$$35\,000$ for decahedral
and $N$$\approx$$80\,000$ for fcc structures if the cluster was constrained to maintain the 
correct point group symmetry ($D_{5h}$ for decahedral and $O_h$ for fcc).
The energies of the optimal clusters for each structural type are shown in Fig.~\ref{fig:LJphased}(a).
The lobed shape of the decahedral and fcc lines results from changes in the shape of the optimal
sequence; as the size increases 
the forms depicted in Fig.~\ref{fig:forms}(a) and (c) become more rounded by the introduction
of $\{110\}$ facets of increasing size, 
and the grooves in the Marks decahedra become deeper.
The energetic crossover sizes are consistent with 
previous results \cite{Raoult89a}:
$N_{\rm icos\rightarrow deca}(T$=0)$\approx$1450 and $N_{\rm deca\rightarrow fcc}(T$=0)$\approx$$213\,000$.

In order to calculate $\overline{\nu}$ the Hessian matrix must be diagonalized, 
and so the sizes for which $\overline{\nu}$ can be calculated are limited to $N$$<$3500. Above this size
we rely upon the extrapolation of Eq.~(\ref{eq:nuvN}).
To construct the melting line $a_m$ is assigned the value of the zero pressure 
bulk melting temperature \cite{vanderHoef00} and the 
other three parameters in Eq.~(\ref{eq:Tm}) are fitted using the melting points
of the first four Mackay icosahedra, which were obtained from Monte Carlo simulations. 

The structural phase diagram that results from these calculations is shown in Fig.~\ref{fig:LJphased}(b).
When interpreting these diagrams, 
one should remember that the phase boundaries divide the plane into regions where the
majority of clusters have a particular equilibrium structure.
If the non-monotonic variation of cluster properties, as illustrated by the examples in Table \ref{table:Tss}, 
were fully taken into account, the phase boundaries would be considerably rougher.

The effect of the vibrational entropy can be clearly seen from the slopes of the phase boundaries in 
Fig.~\ref{fig:LJphased}(b),
e.g.\ $N_{\rm icos\rightarrow deca}(T_m)$$\approx$6550 and $N_{\rm deca\rightarrow fcc}(T_m)$$\approx$$10\,600\,000$.
At higher temperatures icosahedra and Marks decahedra remain most stable up to considerably larger sizes 
This is because of the relative values of their vibrational frequencies: 
$a_{\nu}^{\rm icos}$ and $a_{\nu}^{\rm deca}$ are 2.06\% and 0.19\% less than $a_{\nu}^{\rm fcc}$,
respectively. Even though the difference between fcc and decahedral frequencies is much smaller,
the effect on the slope of the phase boundary is larger because of the larger value of $N$ in
the denominator of Eq.~(\ref{eq:t_ss_est}).

\begin{figure}
\begin{center}
\includegraphics[width=8.2cm]{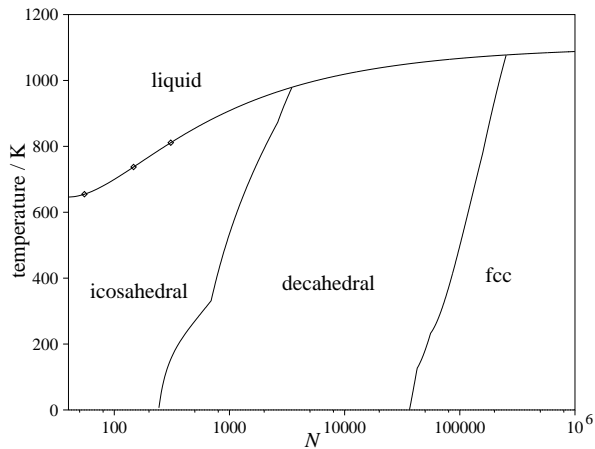}
\caption{\label{fig:SCphased}
Structural phase diagram for silver clusters.
}
\end{center}
\end{figure}

The techniques that we have developed are readily applicable to other atomic clusters. 
This is illustrated in Fig.~\ref{fig:SCphased} for silver clusters 
described by the Sutton-Chen potential \cite{Sutton90}. 
The phase diagram has a very similar form to the LJ diagram except that 
the crossovers occur at smaller sizes, and again the icosahedra and decahedra are
substantially stabilized by temperature.
$N_{\rm icos\rightarrow deca}(0)$$\approx$240 and $N_{\rm deca\rightarrow fcc}$(0)$\approx$$36\,600$ compared to 
$N_{\rm icos\rightarrow deca}(T_m)$$\approx$3500 and $N_{\rm deca\rightarrow fcc}(T_m)$$\approx$$253\,000$.

However, the method cannot be simply applied to molecular clusters because 
the orientational degrees of freedom add an extra degree of complexity.
Tests on N$_2$ clusters showed that for minima where the molecular centres have a common structure but 
the molecules have different orientations there is a considerable variation in $\overline{\nu}$.
Thus one of the assumptions underlying the use of Eq.~(\ref{eq:t_ss_est}) is invalidated and
one would have to consider the competition between orientational isomers
as well as between structural types.  

The key role that we have shown the temperature to play in determining 
the equilibrium structure of a cluster underlines the importance of only comparing
clusters that have both the same size and temperature. 
This may help to explain some of the apparent contradictions between 
experiments and zero-temperature theoretical calculations and 
between different experiments (e.g.\ electron diffraction \cite{Farges88} 
and EXAFS \cite{Kakar97} predict different crossover sizes for rare gas clusters).

Although the developments in this paper allow a more complete comparison of 
theory and experiment, it should be remembered that equilibrium may be hard 
to achieve in experiment. For example, experiments on gold clusters \cite{Andres96} and clusters of C$_{60}$
molecules \cite{Branz00} show the importance of annealing the clusters at 
sufficiently high temperature to locate the most stable forms. Furthermore,
the (free) energy barrriers between structural types \cite{Doye99c,Doye99f}, which have been 
shown to be large for LJ clusters, are likely to increase with size, making 
structural transformations increasingly hard.
In such a situation, growth may preserve the dominant structure at the last size at which
equilibrium was obtained \cite{Baletto00}, or may occur around a core structure 
that allows rapid growth \cite{van96b}.

JPKD is grateful to Emmanuel College, Cambridge for financial support,
and FC has been supported by a European Community Marie Curie Fellowship.

\end{document}